\documentclass[aps,pre,twocolumn,groupedaddress,showpacs,showkeys]{revtex4}

\usepackage{graphicx}% Include figure files
\usepackage{dcolumn}% Align table columns on decimal point
\usepackage{bm}% bold math
\usepackage{amssymb}

\begin{document}

\title{Detecting synchronization of self-sustained oscillators by
external driving with varying frequency\footnote{\large\sf
Published in Phys. Rev. E 73, 026208 (2006)}}

\author{Alexander~E.~Hramov}
\email{aeh@cas.ssu.runnet.ru}
\author{Alexey~A.~Koronovskii}
\email{alkor@cas.ssu.runnet.ru} \affiliation{Faculty of Nonlinear
Processes, Saratov State University, Astrakhanskaya, 83, Saratov,
410012, Russia}
\author{Vladimir I. Ponomarenko}
\email{vip@sgu.ru}
\author{Mikhail D. Prokhorov}
\affiliation{Saratov Department of the Institute of
RadioEngineering and Electronics of Russian Academy of Sciences,
Zelyonaya, 38, Saratov, 410019, Russia}

\date{\today}

\begin{abstract}
We propose a method for detecting the presence of synchronization
of self-sustained oscillator by external driving with linearly
varying frequency. {The method is based on a continuous wavelet
transform of the signals of self-sustained oscillator and external
force and allows one to distinguish the case of true
synchronization from the case of spurious synchronization caused
by linear mixing of the signals.} We apply the method to driven
van der Pol oscillator and to experimental data of human heart
rate variability and respiration.
\end{abstract}


\maketitle

\vskip 0.5cm

\section{Introduction}
\label{Sct:Introduction}

It is well  known that interaction between nonlinear oscillatory
systems including the ones demonstrating chaotic behavior can
result in their synchronization. Various types of synchronization
between oscillatory processes have been intensively studied in
many physical, chemical, and biological systems
\cite{Pikovsky:2002_SynhroBook, Boccaletti:2002_ChaosSynchro,
Glass:1989_BioChaosBook, Glass:2001_SynchroBio,
Mosekilde:2002_BioSynchroBook, Schafer:1999_cardio}. Of particular
interest in recent years is the investigation of synchronization
in living organisms whose activity is determined by interaction of
a great number of complex rhythmic processes. The underlying
sources of these oscillatory processes often cannot be measured
separately, but only superpositions of their signals are
accessible. For example, in electroencephalography recordings on
the scalp the measured signals are the superpositions of the
signals generated by various interacting sources. {As a result,
one can detect spurious synchronization between brain sources and
come to wrong biological conclusions
\cite{Meinecke:2005_Prosachivanie}. Such situation is typical for
many multichannel measuring devices. One faces the similar problem
studying synchronization between the rhythms of the cardiovascular
system (CVS)}. The most significant oscillating processes
governing the cardiovascular dynamics, namely, the main heart
rhythm, respiration, and the process of blood pressure slow
regulation with {the} fundamental frequency close to 0.1 Hz,
appear in various signals: electrocardiogram (ECG), blood
pressure, blood flow, and heart rate variability (HRV)
\cite{Stefanovska:2000_cardio}. This fact impedes studying their
synchronization.

Synchronization between the main rhythmic processes in the human
CVS has been reported in Refs. \cite{Schafer:1999_cardio,
Stefanovska:2000_cardio_Physica_A, Janson:2002_cardio,
Prokhorov:2003_HumanSynchroPRE, Rzeczinski:2002_cardio}. It has
been found that the systems generating the main heart rhythm and
the rhythm of slow regulation of blood pressure can be treated as
self-sustained oscillators and the respiration can be regarded as
an external forcing of these systems
\cite{Prokhorov:2003_HumanSynchroPRE, Rzeczinski:2002_cardio}.
{However, at respiration frequencies close to 0.1~Hz it becomes
difficult to distinguish the case of true synchronization between
the respiration and the process of blood pressure regulation from
the case of spurious synchronization caused by the presence of the
respiratory component in the HRV and blood pressure signals used
for the analysis of the rhythm with the basic frequency of about
0.1~Hz.} Actually, the presence of external forcing can result in
linear mixing of the driving signal and the signal of the
self-sustained oscillator without any synchronization. Another
possible case is the simultaneous presence of mixing of the
signals and their synchronization.

{In this paper we propose a method for detecting synchronization
of self- sustained oscillator by external driving with linearly
varying frequency. The method is based on a wavelet transform of
both the external signal and the self-sustained oscillator signal
and allows one to distinguish the true synchronization from the
spurious one caused by linear mixing of the signals. The principal
moment of our approach is that we vary the frequency of external
signal. We verify our method applying it to driven van der Pol
oscillator and to experimental data of human HRV and respiration.
It should be noted that synchronization between the different
rhythmic processes can be detected {also} from the analysis of
univariate data \cite{Janson:2002_cardio, AddRefNumber1,
AddRefNumber2, Jamsek:2003_Bispectral, Jamsek:2004_Bispectral2,
Ponomarenko_2005_CSF}. However, in the present paper we use only
bivariate signals for detecting synchronization}.

The paper is organized as follows. In Sec.~II the models for
studying the effects of synchronization and mixing are considered.
In Sec.~III we describe the method for detecting synchronization
{using the continuous} wavelet transform. Section~IV presents the
results of method application to driven asymmetric van der Pol
oscillator. In Sec.~V the synchronization between the respiration
and the process {of slow regulation of blood pressure} is studied.
In Sec.~VI we summarize our results.

\section{Models}
\label{Sct:Model}

{The universality of the phenomena observed in periodically driven
self-sustained oscillators of physical and physiological nature
has been discussed in Refs. \cite{Schafer:1999_cardio,
Janson:2002_cardio}. It has been shown there that qualitatively
the same features of synchronization are observed in the case of
periodic driving of van der Pol oscillator and in the case of
respiratory forcing of the heartbeat and the process with the
basic frequency of about 0.1~Hz. Let us consider the asymmetric
van der Pol oscillator under external forcing with linearly
increasing frequency as a model for studying interaction between
the respiration and the process of blood pressure slow
regulation}.

The periodically driven asymmetric van der Pol oscillator is
described by the following equation:
\begin{equation}
\ddot{x}-\mu\left(1-\alpha
x-x^2\right)\dot{x}+\Omega^2x=K\sin\varphi(t), \label{eq:VdP}
\end{equation}
where $\mu=1$ is the parameter of nonlinearity, $\Omega=0.24\pi$
is the natural frequency, and $K$ and $\varphi(t)$ are,
respectively, the amplitude and phase of the external force. The
phase
\begin{equation}
\varphi(t)=2\pi\left[(a+bt/T)\right]t \label{eq:ExtSignalPHASE}
\end{equation}
defines the linear dependence of the driving frequency $\omega_d(t)$ on
time:
\begin{equation}
\omega_d(t)=\frac{d\varphi(t)}{dt}=2\pi(a+2bt/T),
\label{eq:ExtSignal}
\end{equation}
{where $a=0.03$, $b=0.17$, and $T=1800$ is the maximal time of
computation. We choose these parameter values to compare the
results of simulation with those obtained at investigation of
experimental signals of respiration and HRV (see
Sec.~\ref{Sct:SSS})}.

The case $\alpha=0$ in Eq.~(\ref{eq:VdP}) corresponds  to the
classical van der Pol oscillator with symmetric limit cycle. As a
result of phase portrait symmetry, the power spectrum of
oscillations has only odd harmonics $(2n+1)f_0$, $n=1,2,\dots$, of
the basic frequency $f_0$. Since the second harmonic $2f_0$ of the
process with the basic frequency {close to $0.1$~Hz} is well
pronounced in the power spectrum, we consider the modified van der
Pol oscillator (\ref{eq:VdP}) with $\alpha=1$.

\begin{figure}
\centerline{\scalebox{0.4}{\includegraphics{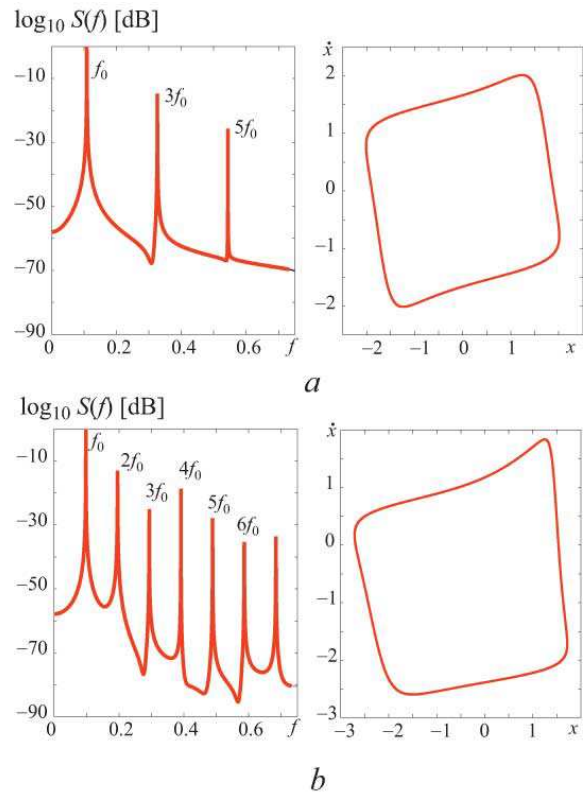}}}
\caption{{(Color online) Power spectra and phase portraits for van
der Pol oscillator (\ref{eq:VdP}) at $K=0$, $\Omega=0.24\pi$, and
$\mu=1$ (a) $\alpha=0$, (b) $\alpha=1$ \label{spectraVdP}}}
\end{figure}

{The difference between the frequency $\omega_0$ of self-sustained
oscillations and the natural frequency $\Omega$ is due to the
effect of nonlinearity ($f_0=\omega_0/2\pi=0.106$ in
Fig.~\ref{spectraVdP}{\it a}). In the case of asymmetric van der
Pol oscillator this difference is greater ($f_0=0.098$ in
Fig.~\ref{spectraVdP}{\it b})}.

To compare the case of synchronization of oscillations  by
external driving with the case of mixing of the signals we
consider the superposition {of the signals}
\begin{equation}
x_\Sigma(t)=x(t)+R\sin\varphi(t),
\label{eq:SumSignal}
\end{equation}
where $x(t)$ is the signal of the autonomous asymmetric van der Pol
oscillator and $R\sin\varphi(t)$ is the additive signal with the
amplitude $R$, the phase $\varphi(t)$, and varying frequency
(\ref{eq:ExtSignal}).

\section{Method of detecting synchronization using continuous wavelet
transform. Measure of synchronization of oscillations}
\label{Sct:WVLTS}

{Studying} synchronization of chaotic oscillators  various
definitions of synchronization have been introduced, namely, the
complete synchronization, generalized synchronization, lag
synchronization, and phase synchronization
\cite{Pikovsky:2002_SynhroBook}. To investigate phase
synchronization one has to choose the way of phase definition for
the chaotic signals \cite{Rosenblum:1996_PhaseSynchro,
Pikovsky:2002_SynhroBook, Anishchenko:2002_SynchroEng,
Pikovsky:2000_SynchroReview}.

We will use the recently proposed approach
\cite{Koronovskii:2004_JETPLettersEngl, Hramov:2004_Chaos,
Aeh:2005_SpectralComponents, Aeh:2005_PhysicaD} to the analysis of
synchronization based on examination of a continuous set of phases
defined with the help of the continuous wavelet transform
\cite{WaveletsInPhysics:2004, alkor:2003_WVTBookEng}
\begin{equation}
W(s,t_0)=\int_{-\infty}^{+\infty}x(t)\psi^*_{s,t_0}(t)\,dt
\label{eq:WvtTrans}
\end{equation}
of the signal $x(t)$, where $\psi_{s,t_0}(t)$ is the wavelet function
related to the mother-wavelet $\psi_{0}(t)$ as
\begin{equation}
\psi_{s,t_0}(t)=\frac{1}{\sqrt{s}}\psi_0\left(\frac{t-t_0}{s}\right).
\label{eq:Wvt}
\end{equation}
The time scale $s$ corresponds to the width of the  wavelet
function $\psi_{s,t_0}(t)$, $t_0$ is the shift of the wavelet
along the time axis, and symbol ``$*$'' denotes complex
conjugation.

We use the Morlet-wavelet \cite{Grossman:1984_Morlet}
\begin{equation}
\psi_0(\eta)=({1}/{\sqrt[4]{\pi}})\exp(j\sigma\eta)\exp\left({-
\eta^2}/{2}\right)
\end{equation}
as the mother-wavelet function. The choice of the wavelet
parameter $\sigma=2\pi$ provides the relation $s=1/f$ between the
time scale $s$ of the wavelet transform and the frequency $f$ of
the Fourier transform.

The wavelet spectrum
\begin{equation}
W(s,t_0)=|W(s,t_0)|\exp[j\phi_s(t_0)] \label{eq:WVT_Phase}
\end{equation}
describes the system dynamics for every time scale $s$ at  any
time moment $t_0$. The value of $|W(s,t_0)|$ determines the
presence and intensity of time scale $s$ at the moment of time
$t_0$. At the same time, the phase $\phi_s(t)=\arg\,W(s,t)$ is
naturally defined for every time scale $s$. In other words, it is
possible to describe the behavior of every time scale $s$ using
its phase $\phi_s(t)$.

If in the signals $x_{1,2}(t)$ there is a range of time  scales
$s_1 \le s \le s_2$ for which the phase locking condition
\begin{equation}
|\Delta\phi_s(t)|=|\phi_{s1}(t)-\phi_{s2}(t)|<\mathrm{const}
\label{eq:PhaseLocking}
\end{equation}
is satisfied and the part of the wavelet spectrum energy in  this
range does not vanish
\begin{equation}
E_{snhr}=\int\limits_{s_1}^{s_2}\langle E(s)\rangle\,ds>0,
\label{eq:SynchroEnergy}
\end{equation}
where $\langle{E(s)}\rangle$ is the distribution of integral
energy by time scales defined as
$\langle{E(s)}\rangle=(1/T)\int_{t}^{t+T}|W(s,t_0)|^2\,dt_0 $,
then the time scales $s\in[s_1;s_2]$ are synchronized and the
oscillators are in the regime of time scale synchronization
\cite{Hramov:2004_Chaos}. In Eq.~(\ref{eq:PhaseLocking})
$\phi_{s1,2}(t)$ are the continuous phases of the first and the
second oscillator corresponding to the synchronized time scales
$s\in[s_1;s_2]$.

Using continuous set of time scales $s$ and the  phases associated
with these scales we introduce the quantitative measure of chaotic
synchronization \cite{Koronovskii:2004_JETPLettersEngl,
Hramov:2004_Chaos}:
\begin{equation}
\gamma=\left.\int\limits_{s_1}^{s_2}\langle{E(s)\rangle}\,ds
\right/\int\limits_{0}^{\infty }\langle{E(s)}\rangle\,ds
\label{eq:Gamma}.
\end{equation}
This measure defines the part of the wavelet spectrum energy
fallen into the synchronized time scales. Increase of $\gamma$
from $0$ to $1$ points to the increase of the part of the wavelet
spectrum energy fallen into the synchronous time scales $s$.

\section{Investigation of driven asymmetric van der Pol oscillator}
\label{Sct:VdP}

\subsection{Amplitude dynamics of wavelet spectra of driven oscillator
and superimposed signal}
\label{SubSct:VdP_Ampl}

\begin{figure}
\centerline{\scalebox{0.5}{\includegraphics{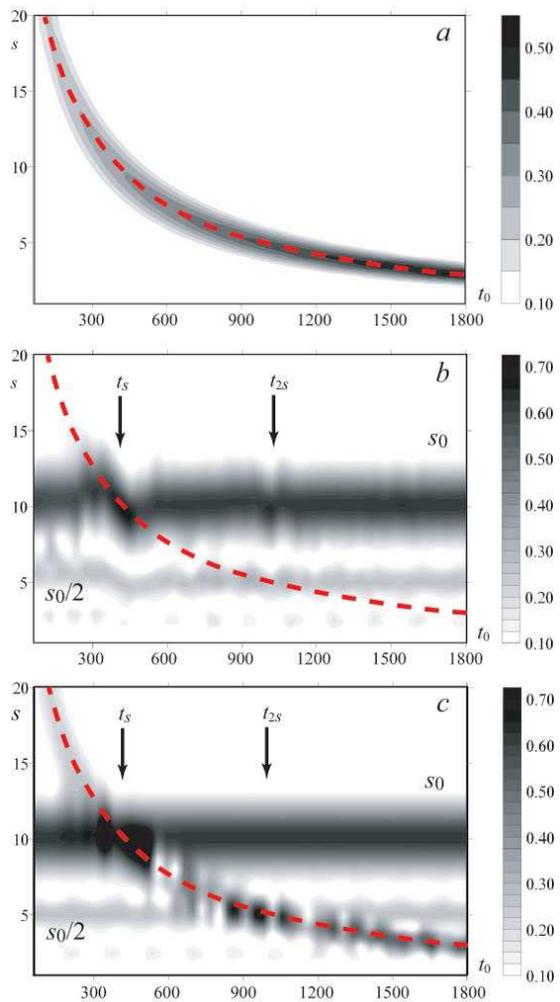}}}
\caption{{(Color online) Shaded plot of the wavelet power spectra
$|W(s,t_0)|$ for the external signal with the frequency varying in
accordance with Eq. (\ref{eq:ExtSignal}) ({\it a}), the signal
generated by oscillator (\ref{eq:VdP}) ({\it b}), and the
superimposed signal (\ref{eq:SumSignal}) ({\it c}). Time is shown
on the abscissa and time scale is shown on the ordinate. The color
intensity is proportional to the absolute value of the wavelet
transform coefficients. The scales from the right side of the
figure show the values of the coefficients. The dashed lines
indicate the time scale $s_d(t)$ corresponding to the linearly
increasing frequency $\omega_d(t)$ (\ref{eq:ExtSignal}}).
\label{Fig:WvltSpectraNonaut}}
\end{figure}

{Let us consider the wavelet power spectra $|W(s,t)|$ of the
external signal $K\sin\varphi(t)$ with linearly varying frequency,
the signal $x(t)$ generated by Eq.~(\ref{eq:VdP}), and the
superimposed signal $x_\Sigma(t)$ (\ref{eq:SumSignal}). Typical
wavelet power spectra of these signals are presented in
Fig.~\ref{Fig:WvltSpectraNonaut}}.

%Предлагаю перенести s0 на рис. (c) в левую часть, как на рис. (b),
%чтобы не путать читателя.

The analysis of the wavelet power spectrum {of the signal $x(t)$
of driven van der Pol oscillator}
(Fig.~\ref{Fig:WvltSpectraNonaut}b) reveals the classical picture
of oscillator frequency locking by the external driving. As a
result of this locking, the breaks appear close to the time
moments $t_s$ and $t_{2s}$ denoted by arrows, when the driving
frequency is close to the frequency of autonomous oscillator or to
its second harmonic. {These breaks represent the entrainment} of
oscillator frequency and its harmonic by external driving. In the
region of the frequency entrainment the amplitude of the
respective coefficients of the wavelet spectrum increases. This
fact agrees well with the known effect of the oscillation
amplitude increase in the synchronization (Arnold) tongue. {If the
detuning $\delta = (\omega_d-2\pi f_0)$ is great enough, the
frequency of oscillations returns to the natural frequency of the
autonomous oscillator.}

It should be noted that besides the break at the main time scale
$s_0$ {of the spectrum}, the break at the scale $s_0/2$
corresponding to the second harmonic $2f_0$ takes place. {In this
region the} wavelet surface has no maxima associated with the
driving signal since the intensity of the corresponding spectral
component is low.

The wavelet  power spectrum of the superimposed signal $x_\Sigma
(t)$ is shown in Fig.~\ref{Fig:WvltSpectraNonaut}c. In contrast to
the previous case, the dynamics of both the van der Pol oscillator
and the external force with varying frequency is well pronounced
in the spectrum. The absence of break in the neighborhood of the
time moment $t_s$ indicates the absence of entrainment of the
oscillator frequency. One can see only slight distortion of the
wavelet surface caused by the increase of the wavelet coefficients
amplitude $|W|$ at the time moments close to $t_s$. This effect is
the result of the addition of two signals with comparable
amplitudes and close frequencies. Furthermore, the surface
distortion in the region of the main scale $s_0$ is not followed
by any change in {the wavelet spectrum at} the scale $s_0/2$
corresponding to the second harmonic of the signal $x(t)$.
Similarly, the coincidence of the frequencies $\omega_d$ and $4\pi
f_0$ at the time moment $t_{2s}$ is not followed by any changes of
the dynamics at the main scale $s_0$.

Thus, it  is possible to distinguish the cases of synchronization
and mixing {of the signals} analyzing the dynamics of time scales
corresponding to the basic frequency and its harmonics in the
wavelet power spectrum. {In the case of mixing the changes in the
dynamics of the scale having the frequency which is close to the
driving frequency, do not lead to any change in the dynamics of
other time scales.} In the case of synchronization the typical
break in the wavelet power spectrum is observed at all
characteristic {time} scales.

\subsection{Phase dynamics of driven oscillator and superimposed
signal}
\label{SubSct:VdP_Phasa}

\begin{figure}
\centerline{\scalebox{0.45}{\includegraphics{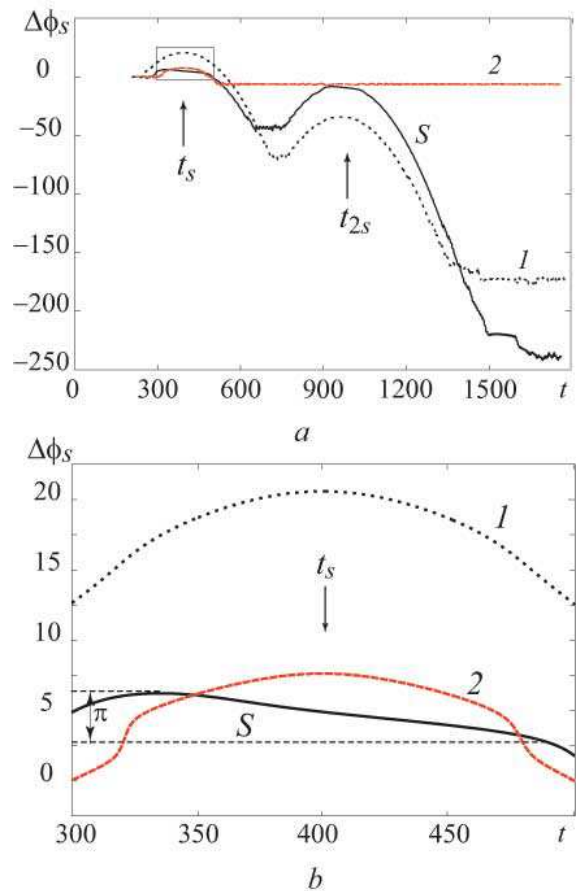}}}
%\centerline{\scalebox{0.55}{\includegraphics{Fig2b.eps}}}
\caption{{(Color online) Phase differences between the
superimposed signal (\ref{eq:SumSignal}) and the external signal
with linearly increasing frequency (curves 1 and 2) and between
the signal of driven van der Pol oscillator (\ref{eq:VdP}) and the
driving signal (curve S). $R=0.2$ for the curve 1, $R=1.0$ for the
curve 2, and $K=0.2$ for the curve S. The phase differences are
calculated at the scale $s_d=2\pi/\omega_d$. The fragment marked
in (a) is enlarged in (b)} \label{Fig:WvltPhasa}}
\end{figure}

{Let us consider the phase difference between the $x(t)$ signal
produced by Eq.~(\ref{eq:VdP}) and the external signal with
linearly increasing frequency and the phase difference between the
$x_\Sigma(t)$ signal produced by Eq.~(\ref{eq:SumSignal}) and the
external signal. The phase differences $\Delta\phi_{s}(t)$ are
calculated along the scale $s_d(t)$, corresponding to the linearly
increasing frequency (\ref{eq:ExtSignal}), i.e., along the dashed
lines in Fig.~\ref{Fig:WvltSpectraNonaut}. Typical dependencies
$\Delta\phi_{s}(t)$ are presented in Fig.~\ref{Fig:WvltPhasa}}.

At first, we consider the case of linear mixing of the signals and
analyze the phase dynamics in the vicinity of $t=t_{s}$, where the
driving frequency is close to the basic frequency of van der Pol
oscillator: $\omega_d(t_{s})\approx 2\pi f_0$. The phase dynamics
at the main scale $s_0=1/f_0$ of van der Pol oscillator can be
written as $\phi_{s0}(t)=2\pi f_0 t+\varphi_0$, where
$\varphi_{0}$ is the initial phase. The phase $\phi_{s}(t)$ of the
driving signal is determined by Eq.~(\ref{eq:ExtSignalPHASE}).
{Under the assumption} that the amplitude of the external signal
is significantly smaller than the amplitude of oscillations of van
der Pol oscillator, the temporal behavior of phase difference
between the components of the superimposed signal $x_\Sigma(t)$
takes the form
\begin{equation}
\Delta\phi_{s}(t)=\phi_{s0}(t)-\phi_{s}(t)=2\pi [(f_0-a)t
-(b/T)t^2)]+\varphi_0. \label{eq:DeltaPhi}
\end{equation}
It follows from Eq.~(\ref{eq:DeltaPhi}) that the phase difference
varies under the parabolic law and the parabola has extremum at
${t=t_{s}}$. In the neighborhood of this moment of time the
dependence $\Delta\phi_{s}(t)$ is symmetric with respect to
$t=t_{s}$. Similar situation takes place at the scale
$s_0/2=1/(2f_0)$ corresponding to the second harmonic.

Such behavior of $\Delta\phi_{s}(t)$ is illustrated by
curves~{\sl1} and {\sl2} in Fig.~\ref{Fig:WvltPhasa}. The
curve~{\sl1} is plotted for small amplitude of the external force
and has a parabolic shape in the neighborhood of $t_s$ and
$t_{2s}$. The curve~{\sl2} plotted for a greater amplitude $R$
also has an extremum at $t=t_s$. However, this extremum is less
pronounced and has a shape different from quadratic one. {It is}
explained by a great influence of the external signal having the
amplitude comparable with the amplitude of the signal generated by
the van der Pol oscillator. This case is characterized by almost
constant phase difference which is close to zero at $t$ far from
$t_s$. For large $R$ values the part of the external signal in the
mixture $x_\Sigma(t)$ is sufficiently large. As a result, the
phase difference is calculated between the phases of practically
the same signals. Therefore, if the part of the driving signal is
large enough in the superimposed signal, one can detect spurious
synchronization using conventional methods of synchronization
investigation.

Let us consider now the case of synchronization of asymmetric van
der Pol oscillator by external driving with linearly increasing
frequency. This case is illustrated by the curve~{\sl S} in
Fig.~\ref{Fig:WvltPhasa}.

We analyze the behavior of the phase difference $\Delta\phi_s(t)$
using the method of slow varying amplitudes under the assumption
that the external driving does not change significantly the
amplitude of nonautonomous oscillations but mainly has the
influence on the phase relationships. It can be shown that
$\Delta\phi_s(t)$ is governed by Adler equation \cite{Adler:1949}
\begin{equation}
\frac{d(\Delta\phi_s(t))}{dt}+\kappa\sin\Delta\phi_s(t)-
\label{eq:Adler}
\end{equation}
$$
-(2\pi f_0-\omega_d(t))=0,
$$
where $\kappa$ is the coefficient defined by the oscillator
parameters. It follows from the Adler equation that in the region
of synchronization defined by the condition $(2\pi
f_0-\omega_d(t))\leq\kappa$ the phase difference $\Delta\phi_s(t)$
monotonically {decreases} by $\pi$ under the driving frequency
variation. By this is meant that in the case of synchronization of
oscillator by external driving with linearly varying frequency we
{will observe} in the vicinity of time moments $t_{s}$ and
$t_{2s}$ the regions where the phase difference varies over $\pi$.
Such situation is illustrated by the curve~{\sl S} in
Fig.~\ref{Fig:WvltPhasa}.

\begin{figure}
\centerline{\scalebox{0.45}{\includegraphics{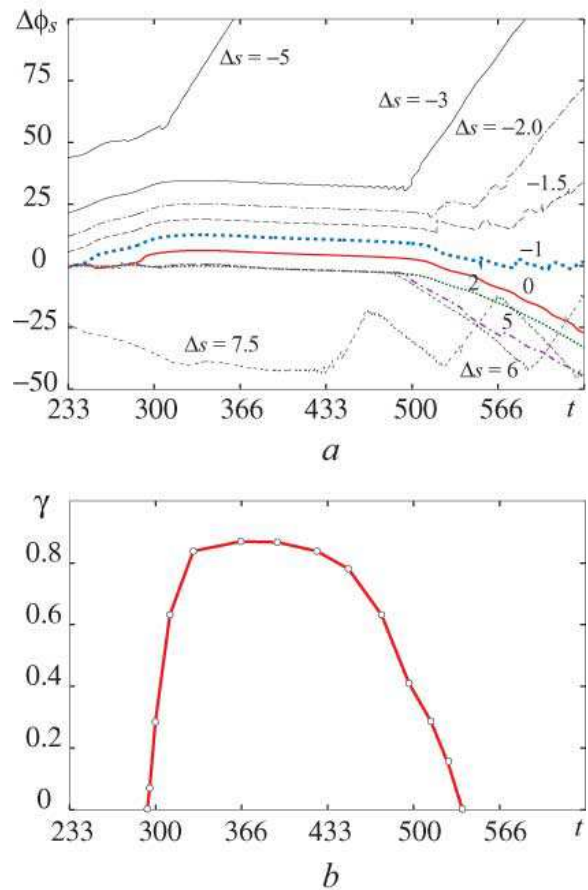}}}
\caption{{(Color online) (a) Phase differences $\Delta\phi_{s}(t)$
between the driven asymmetric van der Pol oscillator and the
driving signal at scales $s_d(t)+\Delta s$. (b) Measure of
synchronization $\gamma$} \label{Fig:WvltPhasaDs_and_Mera}}
\end{figure}

A special feature of our approach for  detecting synchronization
is the application of the external driving with varying frequency.
Another important feature of our method is the analysis of phase
difference between the signals at different time scales $s$
\cite{Hramov:2004_Chaos,Hramov:2005_Chaos_BWO}. We consider the
behavior of phase difference $\Delta\phi_s(t)$ at {the} scales
$s(t)=s_d(t)+\Delta s$, where $\Delta s$ is the detuning of the
scale with respect to the basic scale $s_d(t)$.

In Fig.~\ref{Fig:WvltPhasaDs_and_Mera}a  the phase differences at
various scales are presented for the case when $\omega_0$ is close
to $\omega_d$. It can be seen from the figure that for $\Delta
s\in(-1,2)$ the phase dynamics is qualitatively similar to the
case of accurate adjustment to the basic scale ${s_d}(t)$. At
greater $\Delta s$ values the duration of epochs of synchronous
behavior becomes shorter and beginning from some values of
detuning there is no synchronization at all. The presence of a
range of scales $\Delta s$ within which the synchronous dynamics
is observed allows one to investigate the system behavior without
accurate adjustment of the scale to the basic scale $s_d(t)$
varying in time. It can be useful for the analysis of experimental
data.

Figure \ref{Fig:WvltPhasaDs_and_Mera}b  illustrates the dependence
of measure of synchronization $\gamma$ defined by
Eq.~(\ref{eq:Gamma}) on time. At $t\gtrsim300$ the system quickly
comes to the regime of time scales synchronization. The part of
the wavelet spectrum energy fallen into the synchronized time
scales is more than $80\%$. {With further increasing of time the
value of $\gamma$} decreases and at $t>540$ it becomes equal to
zero. The latter case corresponds to the absence of
synchronization.

\subsection{Dynamics of nonautonomous system at simultaneous presence
of mixing and synchronization} \label{SubSct:VdP_Sum2}

\begin{figure}
\centerline{\scalebox{0.4}{\includegraphics{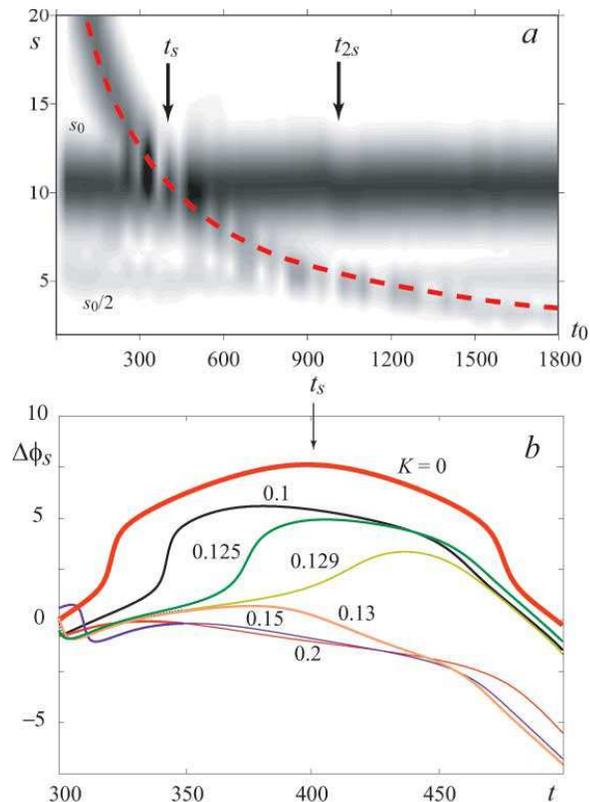}}}
\caption{{(Color online) (a) Wavelet power spectrum  $|W(s,t_0)|$
for the superposition of the signal of driven oscillator
(\ref{eq:VdP}) with $K=0.125$ and the external forcing with $R=1$.
(b) Phase difference at the scale $s_d(t)$ between the external
forcing with $R=1$ and the superposition of the external forcing
and the driven oscillator (\ref{eq:VdP}) signal with different $K$
values} \label{Fig:Sum_and_Prosach}}
\end{figure}

Let us consider {now} the situation when  the effects of mixing
and synchronization of the signals of self-sustained oscillator
and external driving are simultaneously present.
Figure~\ref{Fig:Sum_and_Prosach}a shows the amplitude spectrum
$|W(s,t_0)|$ of the wavelet transform for the signal being the
superposition {of the driven oscillator (\ref{eq:VdP}) signal}
synchronized by external driving with the amplitude $K=0.125$ and
the driving signal itself with $R=1$. This spectrum has the
features inherent to both the cases of synchronization and mixing
and gives no definite answer to the question about the nature of
the underlying dynamics. The analysis of phase dynamics appears to
be more informative.

Figure~\ref{Fig:Sum_and_Prosach}b  shows the dynamics of phase
{differences} $\Delta\phi_{s}(t)$ for $R=1$ and various amplitudes
$K$. At small $K$ values the situation is qualitatively similar to
the considered above case of mixing without synchronization. The
regime of synchronization appears in a certain range of
frequencies as the amplitude $K$ increases. The variation of phase
difference in the synchronization tongue takes the form of an
inclined line. The boundaries of this line correspond to the loss
of synchronization. Note that variation of phase difference in the
synchronization region {in Fig.~\ref{Fig:Sum_and_Prosach}b} is
smaller than in the case of $R=0$.

\section{Investigation of synchronization between the respiration and
the
process of blood pressure slow regulation}
\label{Sct:SSS}

This section  contains the results of physiological data analysis.
We studied eight healthy young male subjects having average levels
of physical activity. The signals of ECG and respiration were
simultaneously recorded in the sitting position with the sampling
frequency 250 Hz and 16-bit resolution. The experiments were
carried out under paced respiration with the breathing frequency
linearly increasing from 0.05~Hz to 0.3~Hz. The rate of breathing
was set by sound pulses. The duration of experiments was 30~min.

\begin{figure}
\centerline{\scalebox{0.7}{\includegraphics{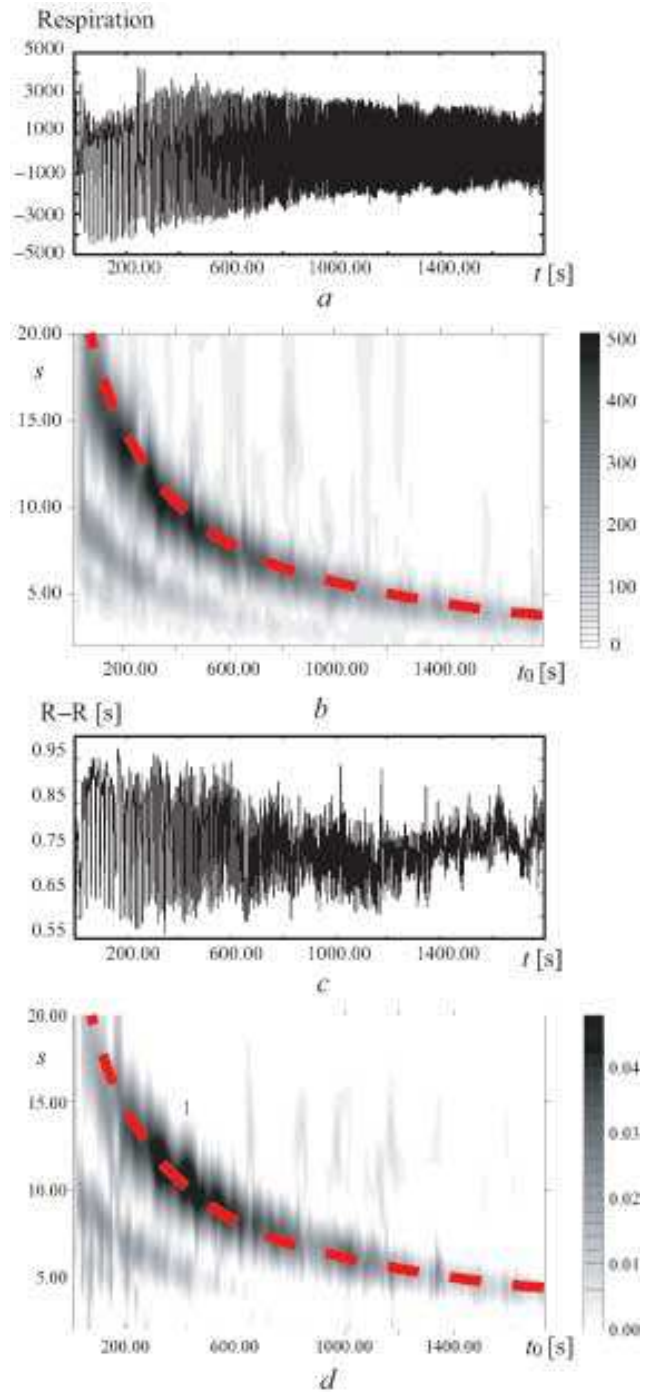}}}
\caption{{(Color online) Typical time series of breathing with
linearly increasing frequency (a) and its wavelet power spectrum
(b). The respiratory signal is in arbitrary units. Sequence of
R--R intervals for the case of respiration with linearly
increasing frequency (c) and its wavelet power spectrum (d). The
dashed lines indicate the time scale corresponding to linearly
increasing frequency of respiration}
\label{Fig:WvltPhasaBespatov}}
\end{figure}

{Figure~\ref{Fig:WvltPhasaBespatov}a shows a typical respiratory
signal with linearly increasing frequency and its wavelet power
spectrum (Fig.~\ref{Fig:WvltPhasaBespatov}b)}. Extracting from the
ECG signals the sequence of R--R intervals, i.e., the series of
the time intervals between the two successive R-peaks, we obtain
the information about the heart rate variability. Typical sequence
of R--R intervals for breathing at linearly increasing frequency
is shown in Fig.~\ref{Fig:WvltPhasaBespatov}c. Since the sequence
of R--R intervals is not equidistant, we developed a technique for
applying continuous wavelet transform to {nonequidistant} data.
Figure~\ref{Fig:WvltPhasaBespatov}d shows the wavelet spectrum
$|W(s,t_0)|$ for the sequence of R--R intervals presented in
{Fig.~\ref{Fig:WvltPhasaBespatov}c. This wavelet spectrum
demonstrates the high-amplitude component denoted as {\sl1}
corresponding to the varying respiratory frequency manifesting
itself in the HRV data.} The second harmonic of the respiration is
observed at a twice-higher frequency. The power of the rhythm with
the basic frequency {of} about 0.1~Hz is rather small and this
frequency is not pronounced in the spectrum. The amplitude of the
respiratory rhythm {in R--R intervals} is much higher than the
amplitude of the rhythm with the frequency 0.1~Hz. {Comparing the
wavelet spectrum of R--R intervals under linearly increasing
frequency of respiration with the wavelet spectra presented in
Secs. III and IV one can come to a conclusion that a significant
mixing of the considered physiological signals takes place without
synchronization.} However, the investigations performed in
Ref.~\cite{Prokhorov:2003_HumanSynchroPRE} with the same
experimental data have clearly shown the presence of 1:1
synchronization between the process of blood pressure slow
regulation and respiration at breathing frequencies close to
0.1~Hz. This synchronization has been observed within the time
interval 200--600~s for each of the subject studied. The wavelet
spectrum in {Fig.~\ref{Fig:WvltPhasaBespatov}d} does not allow to
detect the presence of synchronization. The increase of HRV
amplitude at frequencies close to 0.1~Hz can be regarded only as
an indirect indication of the presence of synchronization.

\begin{figure}
\centerline{\scalebox{0.45}{\includegraphics{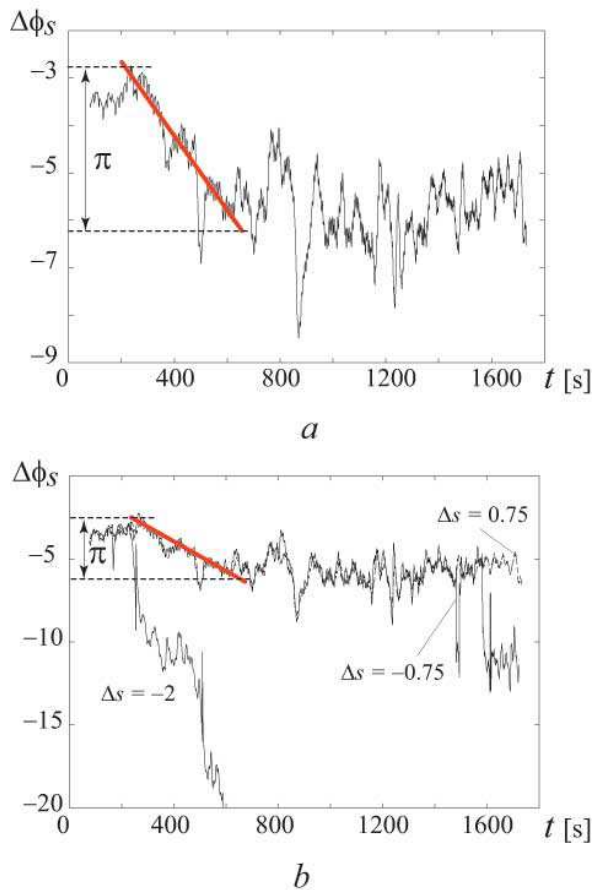}}}

\caption{{(Color online) Phase differences between R--R intervals
and respiratory signal with linearly increasing frequency. The
phase differences are calculated at the linearly varying scale
$s_b(t)=1/f_b(t)$ (a) and scales $s_b(t)+\Delta s$ (b)}
\label{Fig:WvltPhasa_Medical}}
\end{figure}

{In Fig.~\ref{Fig:WvltPhasa_Medical}a the phase difference between
the R--R intervals and respiration is presented. Between 200 and
600~s the phase varies in average almost linearly at the scale
$s=s_b(t)=1/f_b(t)$, where $f_b$ is the linearly increasing
frequency of breathing.} In this time interval the phase variation
is close to $\pi$ indicating the presence of synchronization.
Outside of the synchronization region the phase difference
fluctuates around a constant value. From the result obtained it
may be concluded that the respiratory dynamics manifested in R--R
intervals affects the rhythm of blood pressure regulation within
the time interval 200--600~s. Outside of this interval the
respiratory component is observed in R--R intervals as a result of
mixing of the signals without their interaction.

Similar  behavior of phase difference is observed at neighbor time
scales. Figure~\ref{Fig:WvltPhasa_Medical}b shows phase
differences at {the} scales $s=s_b(t)+\Delta s$. For small
detuning ($\Delta s=\pm0.75$) the phase dynamics is qualitatively
similar to the {dynamics} at the scale $s=s_b$. Hence, a small
error in determining the required scale will not result in
qualitative changes in the plot. {If the mismatch of the scale of
observation and the scale of respiration $s_b$ is large, the
amplitude of the corresponding rhythm decreases and its phase
cannot be defined confidently.} It is the case $\Delta s=-2$ in
Fig.~\ref{Fig:WvltPhasa_Medical}b. As a result, the phase
difference is no more constant.

\section{Conclusion}
\label{Sct:Conclusion}

{We have shown that applying the external driving with varying
frequency to the self-sustained oscillator and using the method
based on the wavelet transform one can distinguish the case of
synchronization of oscillator by this driving signal from the case
of absence of synchronization. The method allows one to detect the
presence of synchronization between the signals with close
frequencies even in the case when the effect of mixing of the
signals is present. The use of the driving signal with varying
frequency is crucial for the proposed method, since calculation of
the phase difference between the superimposed signals at a fixed
time scale gives no answer to the question, whether the system
response is the result of active interaction of the oscillating
processes or the result of their mixing without changes of the
underlying dynamics. The proposed method does not require the
accurate adjustment of the scale of observation to the time scale
associated with the varying frequency of the external driving. The
efficiency of the continuous wavelet transform for estimation of
the signal phase is demonstrated with experimental physiological
data characterized by high level of noise.}

%{The proposed technique is applicable to the situation when having
%at the disposal the superimposed signals it is necessary to define
%whether the system 1 which frequency is adjustable affects on the
%system 2 and can synchronize it.

The proposed technique is applicable to the situation with
superimposed signals being at the disposal, and it is necessary to
define whether system 1 with adjustable frequency affects system 2
and can synchronize it. Such information can be useful for example
for the diagnostics of the cardiovascular system state.
Synchronization between the rhythmic processes in the human
cardiovascular system under paced respiration is less effective in
patients than in healthy subjects, and this effect correlates with
the seriousness of the heart failure \cite{AddRefNumber3}. Using a
small variation of the relative phase or high values of phase
synchronization index as criteria of synchronization one can come
to wrong biological conclusions due to the mixing of analyzed
signals.

The proposed method can be used for detecting synchronization in
{different} systems if it is possible to vary the frequency of the
external driving in the experiment.

\section*{Acknowledgments}
\label{Sct: Acknowledgments}

We thank Dr. Svetlana V. Eremina for English language support and
Dr. Marina V. Hramova for technical support.
This  work is supported by the Russian Foundation for Basic
Research, Grants No.~03--02--17593 and No.~05--02--16273, and U.S.
Civilian Research Development Foundation for the Independent
States of the Former Soviet Union, Grant No.~REC--006. A.E.H.
acknowledges support from CRDF, Grant No.~Y2--P--06--06. A.E.H.
and A.A.K. thank ``Dynasty'' Foundation and ICFPM for financial
support. M.D.P. acknowledges support from INTAS, Grant
No.~03--55--920.

%\bibliography{Aeh}

\end{document}